\documentclass[aps,showpacs,preprintnumbers,amsmath,amssymb,
twocolumn 
]{revtex4}

\usepackage[dvips]{graphicx}
\usepackage{bm}

\begin{document}

\bibliographystyle{apsrev} 

\title {Classical paths for Yang-Mills field with fixed energy}

\author{M. Yu. Kuchiev} \email[Email:]{kmy@phys.unsw.edu.au}

\affiliation{School of Physics, University of New South Wales, Sydney
  2052, Australia}


    \date{\today}

    \begin{abstract} 
A new classical solution for the SU(2) Yang-Mills theory, in which the Euclidean energy plays a role of a parameter is found. A correspondence between this solution and the known selfdual multi-instanton configuration, which has the topological charge $N$, is discussed, the number of parameters governing the new solution is found to be $8N+1$.  For negative energies the new solution is periodic in Euclidean time, for positive energies it exhibits the effect of localization, which states that the solution is completely described within a finite interval of time, for zero energy the found solution is reduced to a selfdual one.
    \end{abstract}


    \maketitle


A number of vital properties of the quantum Yang-Mills field can be studied using the semiclassical approach, which is based on classical solutions. An important class of these solutions presents the BPST instanton of Ref.\cite{Belavin:1975fg} and its generalizations. For a review and summary of the properties of the instanton see e.g. Ref.\cite{Prasad:1980yy}, 
its numerous applications in QCD are outlined in the book \cite{Shuryak:book2003}. A particular way for generalizing the instanton to arbitrary topological charges was discussed in  Refs. \cite{Witten:1976ck,Corrigan:1976vj,Wilczek}, as well as in the 't Hooft unpublished work, see the references in \cite{Prasad:1980yy}. This generalization is often referred to as the CFTW construction. The general formulation of the multi-instanton solution was given by the ADHM construction of Ref.\cite{Atiyah:1978ri}. Another direction for modification of the instanton solution originated in the caloron of Refs. \cite{Harrington:1978ve,Harrington:1978ua}, whose periodic nature made it interesting for applications at finite temperatures. Recent developments and a list of references for the caloron can be found in reviews \cite{vanBaal:2009pn,Weinberg:2006rq}. An interesting  alternative approach to the periodic conditions, called periodic instantons, was suggested in Ref. \cite{Khlebnikov:1991th}. It proved useful for applications related to multiparticle production and studies of the barion and lepton number violation in high energy collisions and early Universe  \cite{Bezrukov:2003qm,Bezrukov:2003er,Rubakov:1996vz}.

It was noted in recent Ref. \cite{Kuchiev:2009rz} that there is an appealing way for generalization of the instanton-type solution, in which the Euclidean energy is considered as a given, fixed parameter. One can foresee that this approach can be useful when the gauge field is exposed to finite temperatures, as discussed in some detail in \cite{Kuchiev:2009rz}, or when reactions with the given energy are considered. 
It was shown in \cite{Kuchiev:2009rz} that a solution with the given Euclidean energy can be constructed using a particular way of `deformation' of the conventional BPST instanton.

The purpose of this work is to construct and study properties of a general classical solution for the Yang-Mills theory, in which the action reaches its minimum provided the Euclidean energy is fixed. Thus, the energy plays a role of the parameter in this solution, which makes it natural 
to refer to this field configuration as the energon. 
It is found that the energon can be obtained by a `deformation' of an arbitrary selfdual solution, which has an arbitrary topological number.

Several properties of the energon resemble properties of the selfdual solution. In particular, the electric and magnetic fields of the energon are collinear, which is in line with the main property of the selfdual solutions, though a ratio of the the fields for the energon is a function of the Euclidean time. When energy in the energon is chosen zero, the energon collapses to the selfdual solution. For arbitrary energies the energon is not selfdual, though it reveals interesting and useful properties, which strongly depend on its energy, as specified below. 


For simplicity we restrict our discussion to the SU(2) gauge theory. The classical action of the gauge field in the Euclidean formulation of this theory reads
\begin{equation}
S  \,=\,\frac{1}{2g^2}\, \int d\tau
\int \! 
\big(\mathbf{E}^a\cdot \mathbf{E}^a+
\mathbf{B}^a\cdot \mathbf{B}^a\big)\,d^3r~.
\label{act}
\end{equation}
Here $\mathbf{E}^a$ and $\mathbf{B}^a$ are the electric and magnetic fields, the superscript $a=1,2,3$ marks isotopic indexes,  the dot-product refers to the scalar product of 3D vectors in the coordinate space. The bold-face notation for the 3D vectors has a conventional meaning
\begin{align}
&\mathbf{E}^a\,\equiv \,E^a_m \,=\,F^a_{m 4}~,
\label{EF}
\\
&\mathbf{B}^a\,\equiv \,B^a_m \,=\,\frac{1}{2}\,\epsilon_{mnl}\,F^a_{nl}~.
\label{BF}
\end{align}
Here $F^a_{\mu\nu}$ are the fields, their subscripts are the indexes of the Euclidean 4D space $\mu,\nu~\text{etc}=1,\dots 4$,  their spatial components are marked by $m,n~\text{etc}=1,2,3$, and $\mu=4$ refers to the temporal component of the Euclidean space.  The sign plus in front of the term $\propto \mathbf{B}^a\cdot \mathbf{B}^a$ in Eq.(\ref{act}) complies with the Euclidean approach. For coordinates in the Euclidean space the usual notation $x_\mu=( \mathbf{r},\tau)$ is used.

We will need an expression for the Euclidean energy of the field
\begin{align}
&\mathcal{E} \,=\,K-V~,
\label{ene}
\\
&K\,=\,\frac{1}{2g^2}\,
\int\mathbf{E}^a\cdot \mathbf{E}^a \,d^3r~,
\label{K}
\\
&V\,=\,\frac{1}{2g^2}\,
\int\mathbf{B}^a\cdot \mathbf{B}^a\,d^3r~.
\label{V}
\end{align}
Here the term $\propto \mathbf{E}^a\cdot \mathbf{E}^a$ plays the role of the density of the kinetic energy of the gauge field, which allows one to interpret $K$ as an effective kinetic energy of the field, while the term $\propto \mathbf{B}^a\cdot \mathbf{B}^a$ gives the density of the effective potential energy, allowing one to consider $-V$ as the corresponding potential energy of the field. Note again the Euclidean formulation, which justifies the sign $V>0$ and makes the second term in (\ref{ene}) negative.

Let us find a local minimum of the action (\ref{act})  assuming that the energy (\ref{ene}) is fixed. Firstly note that the conventional Cauchy's inequality allows one to write
\begin{equation}
\int \!\mathbf{E}^a\cdot \mathbf{E}^a d^3r\,\int \!\mathbf{B}^a\cdot \mathbf{B}^a
d^3r\,\ge\, \left(\int\! \mathbf{E}^a\cdot \mathbf{B}^a\,d^3r \right)^2\!.
\label{cauchy}
\end{equation}
This implies that
\begin{align}
\bigg(\int\! \big(\mathbf{E}^a\cdot \mathbf{E}^a 
& + \mathbf{B}^a\cdot \mathbf{B}^a\big)\,d^3r\bigg)^2  \,\ge\,
\label{E>}
\\
&\Big(2g^2 \mathcal{E}  \Big)^2
+4\left(\int\! \mathbf{E}^a\cdot \mathbf{B}^a\,d^3r\right)^2~,
\nonumber
\end{align}
where, remember, $\mathcal{E}  $ is defined in (\ref{ene}) and is presumed a constant. From the latter inequality one derives the following restriction on the action
\begin{equation}
S  \ge\int\!\! d\tau \left[ \big(\mathcal{E}  \big)^2 
+\left(\frac{1}{g^2}  \int\! \mathbf{E}^a\cdot \mathbf{B}^a\,d^3r \right)^{\!2}\,\right]^{1/2}\!\!.
\label{estAc}
\end{equation}
Clearly, the identity here can be reached if, and only if, the equality in Eq.(\ref{cauchy}) is established. The latter takes place provided the electric and magnetic fields are collinear, which means that the equality in (\ref{estAc}) needs that
\begin{equation}
\mathbf{E}^a(x)\,=\,k(\tau)\,\mathbf{B}^a(x)~.
\label{EkB}
\end{equation}
All integrations in Eq.(\ref{cauchy}) run over spatial coordinates $\mathbf{r}$ only. As a result the coefficient $k(\tau)$ in Eq.(\ref{EkB}) is allowed to be an arbitrary real-valued function of $\tau$. 

So far our search for the minimum of the action has resulted  in Eq.(\ref{EkB}), which should be considered as an equation on the potentials. Obviously, it is a non-linear differential equation, in which the potentials appear in linear and second order terms. 
Still this is a simplification compared to the conventional form of the classical equations, which reads $\nabla_\mu F_{\mu\nu}=0$ and includes nonlinear terms of the third order. 

Moreover, Eq.(\ref{EkB}) admits explicit solution. To describe it we start from a simplified form of (\ref{EkB}) by taking the coefficient there as unity, $k(\tau)=1$. In this particular case Eqs.(\ref{EF}),(\ref{BF}) show that (\ref{EkB}) can be written in the known form
\begin{equation}
F^a_{\mu\nu}\,=\,\tilde F^a_{\mu\nu}\,\equiv \,\frac{1}{2}\,\varepsilon_{\mu\nu\lambda\sigma}
\,F^a_{\lambda\sigma}~,
\label{FF}
\end{equation}
where $\varepsilon_{1234}=1$. Obviously, this form represents the condition of selfduality. (Similarly, by taking  $k(\tau)=-1$ one would end up with the anti-selfdual condition.) A general solution of the self-dual condition is given by the known multi-instanton ADHM configuration \cite{Atiyah:1978ri}. To acknowledge this fact we state that if
$k(\tau)=1$, then the solution of Eq.(\ref{FF}) is $F^a_{\mu\nu,\,\text{sd}  }(x)$, where the subscript `sd' is used throughout to indicate a selfdual solution with an arbitrary topological charge.

Let us generalize this result for the case $k(\tau)\ne \pm\,1$. Introduce new coordinates 
\begin{equation}
z_\mu\,=\,(x_1,x_2,x_3,q(\tau))~,
\label{z}
\end{equation}
where $x_4=\tau$, and $z_4=q(\tau)\equiv q$ is some real-valued function, which precise properties reveal themselves below.  
Thus, spatial coordinates of $z_\mu$ are identical to spatial coordinates of $x_\mu$, while their temporal coordinates are different. Define also the potentials $A^a_\mu(x)$ as follows
\begin{align}
&A^a_m(x)\,=\,A^a_{m,\,\text{sd}  }(z)~,
\label{Am}
\\
&A^a_{\,4}\,(x)\,=\,A^a_{\,4,\,\text{sd}  }(z)~\dot{q}~,
\label{A4}
\end{align}
where $\dot q\equiv dq/d\tau$, and $A^a_{\mu,\,\text{sd}  }$ are the potentials of a selfdual solution. One immediately observes that the fields, which correspond to the potential $A^a_\mu(x)$ satisfy
\begin{align}
&\mathbf{B}^{a}(x)\,=\, {\mathbf{B}_\text{sd}  ^{a}}(z)~,
\label{BB}
\\
&\mathbf{E}^{a}(x)\,=\, {\mathbf{E}_\text{sd}  ^{a}}(z)~\dot q(\tau)~.
\label{EE}
\end{align}
They show resemblance with the fields of the instanton, but are different in that the moment of the Euclidean time for these new fields equals $\tau$, while the temporal coordinate of the selfdual fields is $q(\tau)$. In addition, a scaling factor $\dot q=dq/dt$  appears in the relation for the electric fields.

The electric and magnetic fields of the instanton solution are equal, 
${\mathbf{E}_\text{sd}  ^{a}}(z)={\mathbf{B}_\text{sd}  ^{a}}(z)$, compare  Eq.(\ref{FF}).
Hence Eqs.(\ref{BB}),(\ref{EE})  imply
\begin{equation}
\mathbf{E}^{a}(x)\,=~ \dot q(\tau)~{\mathbf{B}^{a}}(x)~.
\label{EB}
\end{equation}
Observe now that this equality coincides with Eq.(\ref{EkB}) provided  the coefficients in these two equations are identified. We conclude that the potential in Eqs.(\ref{Am}),(\ref{A4})  is a solution of Eq.(\ref{EkB}) provided $k(\tau)=\dot q(\tau)$.

The found solution incorporates the function $q(\tau)$, which up to this point has been treated as arbitrary. Let us specify this function. Substitute Eqs.(\ref{BB}),(\ref{EE}) into (\ref{ene}), presenting the later one as follows
\begin{equation}
\mathcal{E}  \,=\,(\dot q^2-1)\,V_\text{sd}  (q)~.
\label{EqK}
\end{equation}
Here 
\begin{equation}
V_\text{sd}  (q)\,=\,\frac{1}{2g^2}\,\int\mathbf{B}^a_\text{sd}  (z)\cdot \mathbf{B}^a_\text{sd}  (z) \,d^3r~,
\label{KEE}
\end{equation}
is, up to the sign, the potential energy of the selfdual solution, compare definitions in Eqs.(\ref{ene}),(\ref{V}). The argument $z$ in the integrand in (\ref{KEE}), which was defined in (\ref{z}), includes $q$, which makes the integral in Eq.(\ref{KEE}) a well defined function of $q$  for a given self-dual solution. Hence we can treat $V_\text{sd}  (q)$ as a known function $q$. Consequently, Eq.(\ref{EqK}) represents a differential equation on $q(\tau)$. After integration it reads
\begin{equation}
\tau\,=\,\int^q_0\frac{dq}{\dot q}~,
\label{dt}
\end{equation}
where $\dot q$ should be treated as a function of $q$
\begin{equation}
\dot q\,=\,\pm\,\left(1+\mathcal{E}  /\,V_\text{sd}  (q)\right)^{1/2}~.
\label{dotqq}
\end{equation}
Correspondingly, the right-hand side in Eq.(\ref{dt}) is a function of $q$ as well. Inversing it one finds the desired function $q=q(\tau)$. We conclude that for a given selfdual solution the function $q(\tau)$ is well defined.

Summing up this discussion, we see that in order to satisfy Eq.(\ref{EkB}) one takes any selfdual solution and deforms it according to Eqs.(\ref{Am}),(\ref{A4}). The function $q(\tau)$, which is used in the process of this deformation is specified by the same selfdual solution, as Eqs. (\ref{dt}),(\ref{dotqq}) show. The result describes a local minimum of the classical action, in which  the Euclidean energy plays the role of a parameter.  As was mentioned, it is convenient to call this field configuration the energon. Thus, by this construction the energon is related to some instanton solution. To acknowledge this fact we will call the selfdual solution in question the `underlying' one, while the energon would be referred to a `descender' of this underlying selfdual solution. 

A general selfdual solution with the topological charge $N$ for the SU(2) gauge group is described by the ADHM construction and is governed by $8N$ parameters,which can be interpreted as orientations, sizes, and locations of the related instantons. We derive from this that the total number of parameters that govern the energon descending from such selfdual solution equals $8N+1$, where one additional parameter is the energy.

It is instructive to write the action $S$ for the energon explicitly. Since energy of the energon is conserved it is natural to presume that the integration over $\tau$ in Eq.(\ref{act}) is taken within finite limits, and introduce the abbreviated action $S_{0}$,
\begin{equation}
S_{0}\,=\,S + \mathcal{E}  \tau~.
\label{S0}
\end{equation}
(The signs agree with the Euclidean formulation.) Using Eqs.(\ref{EB}),(\ref{EqK}) one derives from Eq.(\ref{act}) that the abbreviated action for the energon reads
\begin{equation}
S_{0}=2 \!\int \dot q^2\,V_{\text{sd}  }(q)\,d\tau=
2\!\int {\dot q}\,V_\text{sd}  (q)\,dq=\!\int \!p\, dq\,.
\label{S0en}
\end{equation}
The last identity here refers to a conventional presentation of the abbreviated action vie the canonical momentum $p$, which is conjugated to the variable $q$. One derives from this that the momentum is equal to
\begin{equation}
p=2 \, \dot q\,V_{\text{sd}  }(q)~.
\label{p}
\end{equation}
Here $\dot q$ is a function of $q$ specified in Eq.(\ref{dotqq}). 

Consider how properties of the energon depend on its energy. Observe that according to Eq.(\ref{dotqq}) the allowed values of energy are restricted from below
\begin{equation}
\,-\max_\tau \big(\, V_\text{sd}  (\tau)\,\big)  \,\le\, \mathcal{E}   \,<\,\infty~.
\label{le}
\end{equation} 
Consider firstly the region of negative Euclidean energies $\mathcal{E}  <0$.  For a selfdual solution with finite topological charge the magnetic field is localized mainly within the finite volume of the Euclidean 4D space.  Hence, Eq.(\ref{KEE}) implies that $V_\text{sd}  (q)\rightarrow 0$ when $q\rightarrow \infty$ and,  consequently, Eq.(\ref{dotqq}) shows that $q$ is bound by condition
\begin{equation}
-V_\text{sd}  (q)\,\le\,\mathcal{E}  <0~,
\label{V<}
\end{equation}
\begin{figure}[tbh]
\centering
\includegraphics[height=5.4 cm,keepaspectratio = true, 
]{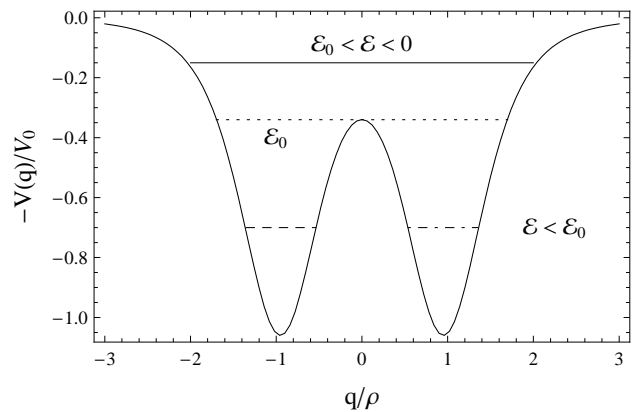}
\caption{
 \label{ras} 
The potential energy of the energon  versus the temporal coordinate $q$ (energy scaled by $V_0=3\pi^2/(\rho g^2)$). The underlying selfdual solution describes two identically oriented instantons of equal size $\rho$ and separation $2\rho$ along the temporal axes. The solid curve shows $-V(q)$ from Eq.(\ref{KEE}). Horizontal lines - Euclidean energies $\mathcal{E}  $: thin line - unique interval of $q$ that satisfies Eq.(\ref{V<}), 
dashed and dot-dashed lines - two different intervals of $q$ that satisfy  Eq.(\ref{V<}), dotted line - a boundary that separates the two regions of $\mathcal {E}$.}
 \end{figure}
\noindent
which specifies one or several finite intervals of $q$. Fig.\ref{ras} illustrates this statement.  It shows a situation, when the underlying selfdual configuration is described by the CFTW multi-instanton construction. In this particular example the two-instanton solution is taken in which two instantons have same size $\rho$ and are separated by the distance $2\rho$ along the temporal coordinate. The figure shows the potential energy $-V(q)$, which is calculated from Eq.(\ref{KEE}) and is depicted versus the coordinate $q$. (Fig. \ref{dva} discussed below presents $-V(q(\tau))$ versus $\tau$.) The potential energy exhibits two minima, one for each of the instantons. The value of the potential energy at the minima is close to the minimum of the potential energy of a single instanton of the same size $\rho$. The later minimum, which equals $-V_0=-3\pi^2/(g^2\rho)$, is chosen as a scaling factor for the energies presented in the figure, while the coordinate $q$ is scaled by the size $\rho$.

One observes that there is a region of Euclidean energies, 
where the inequality in Eq.(\ref{V<}) is satisfied within only one interval of $q$, see the energy level shown by a thin horizontal line in Fig. \ref{ras}. 
For sufficiently low energies
there are two intervals of $q$ where Eq.(\ref{V<}) is satisfied, each located within the limits set by a shape of the corresponding  minimum of $-V(q)$, as the dashed and dot-dashed lines show.

From Eq.(\ref{dotqq}) one derives that for each interval of $q$, which 
satisfies Eq.(\ref{V<}) for a given Euclidean energy, there exists a solution of (\ref{dotqq}) described by a periodic function $q(\tau)$, which 
oscillates within the limits provided by the given interval.
Each such function describes the particular energon. Therefore if for some energy several functions $q(\tau)$ are present, there are several corresponding energons.

For a particular value of $\tau$, when $q$ reaches the boundary of the interval, the function $q(\tau)$ satisfies $\dot q=0$. Eq.(\ref{EB}) shows that at this point the electric field is absent. Everywhere else $\dot q\ne 0$ and the electric field is not zero. However, Eq.(\ref{dotqq}) ensures a restriction from above, $|\dot q|< 1$, which according to Eq.(\ref{EB}) shows that each component of the electric field is necessarily smaller than the corresponding magnetic one  
\begin{equation}
|{E}^{a}_m| <|{B}^{a}_m|~, \quad\quad \quad \mathcal{E}  <0~.
\label{E<B}
\end{equation}
When the Euclidean energy is close to the local minimum of the function $-V(q)$, the electric field of the energon is small. Eventually, when  the energy converges to the minimum, the electric field disappears.  Correspondingly, the kinetic energy of the energon at this point is zero,  while its potential energy equals the minimum of the potential energy of the underlying instanton solution. The example presented in Fig.\ref{ras} shows two degenerate minima of the potential energy (the degeneracy is due to the fact that the instantons are chosen of identical size), so that  when $\mathcal{E}$ reaches the minimum of the potential energy the energon can be located at any of the two available minima, having there zero kinetic energy.

Previously a situation of this type has been encountered in Ref. \cite{Ostrovsky:2002cg} in relation to the sphaleron problem. Remember that the sphaleron was firstly considered in Ref.\cite{Harrington:1978ve,Harrington:1978ua} as a self-consistent solution for interacting gauge and Higgs fields. Ref. \cite{Ostrovsky:2002cg} argued that the pure gauge field exhibits the sphaleron-type behavior when it is treated using particular restrictions. 
The authors of this work found a particular configuration of the gauge field, which possesses only the magnetic field. The energon allows one to relate this construction, call it the COS-sphaleron, to the single-instanton problem, as discussed in \cite{Kuchiev:2009rz}. The consideration above generalizes this statement by showing that the energon, which descends from an arbitrary multi-instanton solution possesses only the magnetic field, provided the Euclidean energy is chosen  at a local minimum of the potential energy of the underlying selfdual configuration.


Consider small Euclidean energies. Take first the case of zero energy, $\mathcal{E}  = 0$. Then Eq.(\ref{dotqq}) shows that $q=\pm \tau$, which indicates that  there are two options. For $q=\tau$ we have $z=x$ and Eqs.(\ref{EE}),(\ref{BB}) show that the energon is identical to the underlying selfdual solution. Similarly, when $q=-\tau$ the energon is identical to the anti-selfdual solution, which differs from the initial, underlying selfdual one by the sign of the temporal component of the potential. From this discussion one derives that for small negative Euclidean energy the function $q(\tau)$ exhibits oscillations as a function of the Euclidean time $\tau$. The smaller the absolute value of the energy, the larger the period of these oscillations, and the larger the interval of $q$, which is covered during these oscillations. For half of the period of these oscillations the energon looks mostly similar to the underlying selfdual solution except the two areas near the stopping points $\dot q=0$. At these points the electric field of the energon is zero, while the electric field of the selfdual solution is not. However, the smaller the absolute value of the energy, the less important the vicinities of these points are since they are located at large $q$ and $\tau$ where all the fields are small anyway. Similarly for the next half-period the energon looks akin the anti-selfdual solution. Thus, during one period of oscillations the energon changes its appearance being close in turn to the selfdual and anti-selfdual solution.

Previously this type of behavior was found for the periodic solution constructed in Ref. \cite{Khlebnikov:1991th}, which was called `periodic instantons'. It is described by the periodic in $\tau$ function, which describes a chain of the instantons and antiinstantons, see also \cite{Kuchiev:2009rz}. Here we observe  a more general manifestation of a similar phenomenon, which takes place for an underlying selfdual solution with an arbitrary topological charge. For negative and low Euclidean energies the energon can be approximated by a chain of the underlying selfdual and related anti-selfdual solutions.

Consider now the region of positive Euclidean energies $\mathcal{E}>0$. Eqs.(\ref{dt}),(\ref{dotqq}) in this case show that the function $q(\tau)$ is monotonic. Moreover it is easy to see  that it varies from $-\infty\le q\le \infty$ when $\tau$ runs over some finite interval, say  $-\tau_0/2\le \tau\le \tau_0/2$, where $\tau_0$ 
is defined using  Eq.(\ref{dt}) as follows
\begin{equation}
\tau_0\,=\,\int_{-\infty}^{\infty}
\frac{dq}
{  \big(1+\mathcal{E}/V_\text{sd}  (q) \,\big)^{1/2}}~.
\label{t0}
\end{equation}
The integral here converges well since at large $q$ the function $V_\text{sd}  (q)$  decreases rapidly. To illustrate the latter statement take  the simplest case when the underlying selfdual solution is the single instanton. Then, taking the conventional gauge one has 
\begin{equation}
F_{\mu\nu,\,\text{ins}}^{\,a}(z)\,=\,-4\,\eta_{\mu\nu}^a \,\frac{\rho^2}{(z^2+\rho^2)^2}~,
\label{Finst}
\end{equation}
where $\rho$ is the instanton size, and from Eq.(\ref{KEE}) one finds
\begin{equation}
V_\text{ins}(q)\,=\,\frac{3\pi^2}{g^2}\,\frac{ \rho^4}{(q^2+\rho^2)^{5/2}}~.
\label{Uinst}
\end{equation}
Substituting this function in  Eq.(\ref{t0}) one derives
\begin{equation}
\tau_0\,=\,\rho \!\int_{-\infty}^{\infty}
\frac{dx}
{  \big(1+\epsilon\,(x^2+1)^{5/2} \,\big)^{1/2}}~.
\label{t0ins}
\end{equation}
Here the parameter $\epsilon=\rho g^2\,\mathcal{E}/\,(3\pi^2)$ is introduced. 
For positive energies it is obviously positive $\epsilon>0$, therefore the integral is convergent and gives a sensible definition for $\tau_0$. Similar result holds for the energon descending from any selfdual solution. 

We see that the discussed definition of the energon at the positive energy is valid only within a finite interval of $\tau$, $\tau_0/2\le\tau\le \tau_0/2$. Inside this interval the function $q(\tau)$ is finite, and its derivative is large, $|\dot q|>1$. Eq.(\ref{EB}) indicates therefore that for positive energies each component of the magnetic field is smaller than the corresponding component of the electric one, 
\begin{equation}
|B^a_m|<|E^a_m|~,\quad\quad\quad \mathcal{E} \,>\,0~.
\label{B<E}
\end{equation}
When $\tau$ is close to the boundaries, $\tau\rightarrow \pm\,\tau_0/2$, the function $q(\tau)$ diverges, $|q(\tau)|\rightarrow \infty$. As a result both electric and magnetic fields are decreasing here, $|E^a_m|,|B^a_m|\rightarrow 0$. 
Eq.(\ref{Finst}), in which  $F^a_{\mu\nu,\,\text{1-ins}}\!\rightarrow 0$ for $|q|\rightarrow \infty$, illustrates this decrease when the energon descends from a single instanton.
Thus the energon as a function of $\tau$  is well localized within the interval  $\tau_0/2\le\tau\le \tau_0/2$. 
\begin{figure}[tbh]
\centering
\includegraphics[height=5.4 cm,keepaspectratio = true, 
]{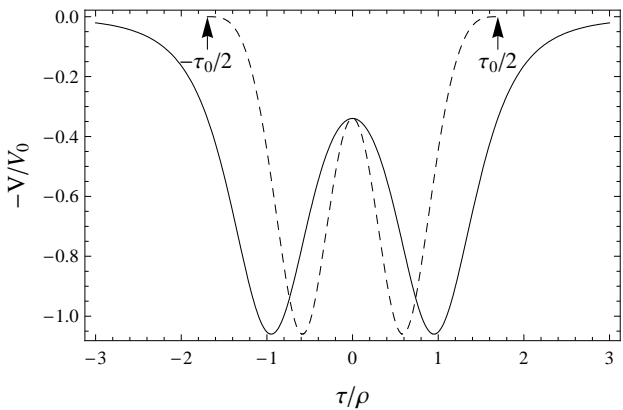}
\caption{
 \label{dva} 
Potential energy for the energon with positive Euclidean energy,  the latter is chosen to be $\mathcal{E}=3\pi^2/(\rho g^2)$. Solid line - potential energy $-V_\text{sd}(\tau)$  vs $\tau$ for the underlying selfdual solution, dashed line - potential energy for the energon  $-V(q(\tau))$ versus $\tau$ calculated from Eq.(\ref{KEE}), in which $q(\tau)$ is defined by Eqs.(\ref{dt}),(\ref{dotqq}). Potential energy and intervals of time are scaled by $V_0=3\pi^2/(\rho g^2)$ and $\rho$ respectively. The arrows show the interval $-\tau_0/2\le \tau\le \tau_0/2$, in which the energon is localized, $\tau_0$ is calculated using Eq.(\ref{t0}). The underlying selfdual solution is the same as the one in the example presented in Fig. \ref{ras}.}
\end{figure}
\noindent

As a slightly more sophisticated illustration consider Fig. \ref{dva}, which shows the potential energy of the energon with positive Euclidean energy. The underlying selfdual solution is taken the same as the one discussed previously in Fig. \ref{ras}. It is a CFTW solution, in which two instantons of the same size $\rho$ are separated by the distance $2\rho$ along the temporal coordinate. The solid line in Fig. \ref{dva} describes the potential energy for this selfdual solution. It has conventional tails at large $|\tau|$. Note the difference in the temporal coordinates, which equals $q$ in Fig. \ref{ras} and $\tau$ in Fig. \ref{dva}. As a result the solid curve in these two figures has a similar shape, but different meaning. In Fig. \ref{ras} it describes the potential energy of the energon, while in Fig. \ref{dva} it represents the potential energy of the underlying selfdual solution. The dashed line in Fig.\ref{dva} presents the potential energy for the energon, which descends from the chosen selfdual solution. The Euclidean energy was taken $\mathcal{E}=3\pi^2/(\rho g^2)$. For this energy and the given selfdual configuration Eq.(\ref{t0}) gives $\tau_0\simeq 3.39 \rho$. Fig. \ref{dva} illustrates the fact that the energon is restricted within the area $-\tau_0/2\le \tau\le \tau_0/2$, which is marked by the arrows. On the boundaries of this region the potential energy is zero, which signals that both the magnetic and electric fields are also zero here in accord with the general properties of the energon for the positive energy.

One can extend the definition for the energon outside the finite interval $-\tau_0/2\le \tau\le \tau_0/2$, but such an extension is not uniquely defined. One can cover the entire axis $-\infty<\tau<\infty$ by an infinite set of finite intervals of time in such a way that
in each interval there exists a particular energon. However, these energons can be defined  differently for different intervals. Firstly, they can be distinguished by the underlying selfdual solution, which can be differently taken for different intervals. Secondly, even if the underlying selfdual solution is chosen the same for all intervals of time, there still remains a freedom to choose the sign in Eq.(\ref{dotqq}), which may be different in different intervals of time. Each energon defined through this procedure is well localized within its own interval of $\tau$, and does not interfere with other energons defined within their own intervals of time. Thus, on the axis $-\infty< \tau<\infty$ there may coexist a variety of different energons. Generically these solutions are not periodic in $\tau$, though
some of them exhibit the periodicity.

Summarizing, this work formulates the classical solution for the Yang-Mills theory in the form of the energon, i.e. solution which minimizes the action provided the Euclidean energy is fixed. An explicit form of the energon is derived from an arbitrary selfdual, multi-instanton solution, which is deformed  using  Eqs.(\ref{Am}),(\ref{A4}). The number of parameters, which govern the energon for the SU(2) gauge group equals $8N+1$, where $N$ is the topological charge of the  multi-instanton solution. The electric and magnetic fields of the energon are collinear,  but in difference with the known property of the selfdual solution they are not equal, see Eq.(\ref{EB}) that includes a function of the Euclidean time, which is defined by the initial selfdual solution via Eq.(\ref{dotqq}). 

The properties of the energon strongly depend on the Euclidean energy. At zero energy there are two options; the energon either equals the underlying selfdual configuration, or coincides with the related anti-selfdual one. For small negative energies there exists one energon, which is related to one periodic function $q(\tau)$ available here. During half of its  period this solution is close to the underlying selfdual configuration, for the other half period to anti-selfdual one. With decrease of energy several energons may appear, each related to a particular periodic function, see Fig. \ref{ras}. For positive energies the energon is localized within a finite interval of the Euclidean time. It can be extended outside this interval, but such an extension is governed by an infinite number of parameters. 

A balance between the electric and magnetic fields also depends on energy. For negative energies each component of the electric field is smaller than the corresponding component of the magnetic field. In particular, when the energy reaches a minimum of the potential energy $V_\text{sd}  (\tau)$, the electric field disappears, and the energon possesses only the magnetic field.  One can call therefore the region of negative energies the `magnetic zone'. 

For positive energies the situation is opposite, the electric field here is always stronger than the magnetic field. The higher the energy, the more strongly 
the electric field prevails over the magnetic one. One can call the region of positive energies the `electric zone'.  Fittingly, the electric ($\mathcal{E}>0$) and magnetic
($\mathcal{E}<0$) zones are separated by the region of zero energy ($\mathcal{E}=0$), where  the energon equals the (anti) selfdual solution whose electric and magnetic fields are equal.


It was mentioned that the selfdual configuration that underlies the energon can be described using the ADHM construction, which has a finite topological charge. Meanwhile, there exist selfdual solutions, which are not described by this construction, for example the caloron. The topological charge of the caloron depends on one's perspective. If, following
\cite{Harrington:1978ve,Harrington:1978ua} one restricts the integration over $\tau$ by the caloron period, then the caloron has the topological charge one. However, if this integration is taken over the whole axis of $\tau$, then the caloron topological charge is infinite, which explains why it cannot be described by the ADHM construction. The point is that the solution found in the present work permits one to consider an arbitrary underlying selfdual configuration, including the caloron or other possible solution with the infinite topological charge. 

There is an appealing way for generalization of the energon. Recall that it is formulated as a classical solution in which the Euclidean energy is fixed, and that  the energy is given by an integral over the 3D space of coordinates $\mathbf{r}$ performed over the particular, quadratic in the gauge fields form. Take now an arbitrary 3D surface $S$ in the 4D Euclidean space, which includes the origin. Consider also an arbitrary quadratic in $\mathbf{E}^a$ and $\mathbf{B}^a$ form $M$ with constant coefficients. Construct a quantity that is given by an integral of this form over the chosen 3D surface, $Q=\int_S M$. 
One can think, for example, that $Q$ is a component of the momentum, or angular momentum of the field, or it can be some quantity that bears a color index. 
Then it seems plausible that the method described in this paper can be modified to allow one to find a classical solution in which $Q$ is presumed fixed. Clearly, if solutions of this type are possible, new interesting opportunities to study properties of the gauge fields would become available.

One can expect that the energon can find applications in the same areas where the selfdual solutions are used. Strong variation of its properties with energy may be useful in various situations. 

\section*{Acknowledgment}

Discussions with Victor Flambaum and Edward Shuryak are appreciated. The work was supported by the Australian Research Council.

\end{document}